# The superfluid density in cuprate high-$T_c$ superconductors – a new paradigm.

J.L. Tallon[1], J.W. Loram[2], J.R. Cooper[2], C. Panagopoulos[2] and C. Bernhard[3].

[1]MacDiarmid Institute for Advanced Materials and Nanotechnology, Victoria University and Industrial Research Ltd., P.O. Box 31310, Lower Hutt, New Zealand.
[2]IRC in Superconductivity, Cambridge University, Cambridge CB3 0HE, England.
[3]Max Planck Institüt fur Festkörperforschung, Heisenbergstrasse 1, D-70506 Stuttgart, Germany.

Abstract

The doping dependence of the superfluid density, $\rho_s \equiv \lambda_{ab}^{-2} \propto n_s/m^*$, of high-$T_c$ superconductors is usually considered in the context of the Uemura relation, namely $T_c$ proportional to $\rho_s$, which is generally assumed to apply in the underdoped regime. We show that a modified plot of $T_c/\Delta_0$ versus $\rho_s$, where $\Delta_0$ is the maximum d-wave gap at $T=0$, exhibits universal features that point to an alternative interpretation of the underlying physics. In the underdoped region this plot exhibits the canonical negative curvature expected when a ground-state correlation competes with superconductivity (SC) by opening up a gap in the normal-state DOS. In particular $\rho_s$ is suppressed much faster than $T_c/\Delta_0$ or indeed $T_c$. The pseudogap is found to strongly modify the SC ground state.

While there is as yet no agreed theory of superconductivity in the HTS cuprates researchers have discerned several systematic trends in their phenomenology. One generally accepted universal property is the proportional relation, $T_c = \text{const} \times \rho_s(0)$, between the superconducting (SC) transition temperature, $T_c$, and the superfluid density, often referred to as the Uemura relation [1-2]. The superfluid density, $\rho_s(T) \equiv \lambda_{ab}(T)^{-2} = \mu_o e^2 \, n_s/m^*$ where $\mu_o$ is the vacuum permeability, e the electronic charge, $n_s$ is the density of superconducting electrons, $m^*$ is their effective mass and $\lambda_{ab}$ is the in-plane London penetration depth. $\rho_s(T)$ is a measure of the phase stiffness of the condensate [3] and plays an important role in governing the irreversibility field [4]. A quasi-linear relation between $T_c$ and $\rho_s$ seems to be broadly applicable to underdoped cuprates irrespective of their structural type and the size of their maximum transition temperature, $T_{c,max}$. As a consequence, the Uemura relation has been extremely influential and is invoked, for example, in support of Bose-Einstein condensation of real-space pairs [2,5], precursor pairing models [3,6], holon condensation in a spin-charge separation model [7] and is generally discussed as a test of theoretical models [8].

Since Uemura's pioneering work [1-2] we have obtained new data which suggest a fundamentally different relation for the doping-dependent trends of $\rho_s(0)$ in HTS. (We do not consider the wider class of materials discussed by Uemura [2]). To set the scene we show in Fig. 1 values of $T_c$ for $Y_{0.8}Ca_{0.2}Ba_2Cu_3O_{7-\delta}$ (Y-123) and $La_{2-x}Sr_xCuO_4$ (La-214) plotted versus $\lambda_0^{-2}$, where the subscript zero refers to $T=0$. The $\rho_s(0)$ data for Y-123 are from transverse-field muon spin relaxation (μSR) studies [9] while the La-214 data are from ac susceptibility and μSR studies [10]. $\rho_s(0)$ data for $Bi_2Sr_2CaCu_2O_{8+\delta}$ (Bi-2212) were obtained from field-dependent thermodynamic measurements [11]. While they show the same trends seen in Fig. 1, they are omitted from the figure for the sake of clarity but will be discussed later in detail. The doping level, expressed as $p$ holes/planar Cu, is just the magnitude of $x$ in $La_{2-x}Sr_xCuO_4$ and, for Y-123 and Bi-2212, is conveniently determined [12] from the approximate empirical relationship $T_c = T_{c,max} [1 - 82.6(p-0.16)^2]$. In each case the maximum $T_c$ in Fig. 1 occurs at *optimal doping*, $p \approx 0.16$, while the maximum superfluid density occurs at *critical doping* [9], $p_{crit} \approx 0.19$. The open symbols in the figure denote the 1/8, optimal and critical doping points.

Fig. 1 reveals several important features. Firstly, the dotted line shows that $T_c \propto \rho_s(0)$ is not sustained, and instead our data exhibit a negative curvature from the lowest doping to $p_{crit}$. In particular, $T_c$ changes little between $p \approx 0.125$ and $p_{crit}$ while $\rho_s$ nearly doubles. It is the origin of this negative curvature below $p_{crit}$ that the present work addresses. Within this general trend, each data set exhibits deviations below 1/8[th] doping which may be associated with incipient stripe instabilities. These anomalies, and a paucity of data points, possibly gave the impression of linearity in earlier data. The behaviour for $p > p_{crit}$ seems more complex since $\rho_s$ falls anomalously [9] in Y-123, Tl-1212 and Bi-2212 (see also ref. [13]) but shows little change in La-214. Leaving aside the overdoped region, it is clear from Fig. 1 that the assumption of linearity is questionable and may conceal important systematic trends.

It is well established that for $p < p_{crit}$ the normal-state (NS) density of states (DOS), $N(E)$, is depleted due to the presence of the pseudogap [14]. The pseudogap energy scale, $E_g$, rises with progressive underdoping and all NS and SC properties are strongly modified either due to the loss of spectral weight or the depletion of states available to scatter into. These effects persist to high temperatures and should not be confused with precursor effects which occur near $T_c$ [15]. For example, the "lost entropy" associated with the pseudogap is not even partially recovered to at least 300K nor the "lost susceptibility" to at least 400K [11]. Many interpretations of the pseudogap have been advanced including, on the one hand, various forms of precursor pairing and, on the other, independent correlations which modify the ground state. Over many years we have argued for the latter [16-18].

In Fig. 2 we compare the doping dependence of $\lambda_0^{-2} \times T_c$ with that of the electronic entropy, $S(T_c)/R$, for Y-123, Bi-2212 and La-214 [11,17]. Here R is the gas constant. With the exception of overdoped La-214 both quantities agree very well across the entire doping range. This correlation is remarkable because it demonstrates a link between a true $T=0$ ground-state property, $\lambda_0$, and a finite-temperature normal-state thermodynamic property, $S(T_c)/T_c$ whose magnitude we know is strongly suppressed by the normal-state pseudogap [11,14]. We conclude then that the superfluid density is also suppressed by the pseudogap and our goal is to determine its quantitative effect on $T_c$ and $\rho_s(0)$.

Assuming the normal-state gap to be non-states-conserving with quasi-triangular energy dependence as indicated by specific heat measurements [18,11], we have solved standard weak-coupling BCS expressions for the normal state DOS:

$$N(E) = N_o \times |E-E_F|/E_g(p) \quad \text{for } |E-E_F| < E_g(p),$$
$$= N_o \quad \text{for } |E-E_F| > E_g(p). \quad (1)$$

Fig. 3 shows $T_c$ plotted as a function of $\zeta = E_g/(2.397 k_B T_c^0)$. The depression in $T_c$ is slow at first and more rapid as $\zeta \to 1$. The superfluid density may be calculated from [19]

$$\rho_s = 1/\Omega \sum [\, 2(\partial \varepsilon_k /\partial k_x)^2 \times \partial f(\Gamma_k)/\partial \Gamma_k + (\partial^2 \varepsilon_k /\partial k_x^2)(1 - \varepsilon_k/\Gamma_k \tanh(\Gamma_k/2k_B T)\,)\,] \quad (2)$$

where $\Gamma_k = \sqrt{(\varepsilon_k^2 + \Delta_k^2)}$. We assume this triangular NS DOS and solve the BCS gap equation to obtain $\rho_s(T)$ from eq. (2). Such a Fermi liquid approach is open to criticism but seems to be valid at lower temperatures where the nodal regions dominate and quasiparticles have long lifetimes. $\rho_s(\zeta)$ is plotted as a function of $\zeta$ in Fig. 3 and is characterised by an initial rapid decrease which slows as $\zeta \to 1$. While details of the pairing model and pseudogap may alter the explicit dependence of $T_c$ and $\rho_s$ on $\zeta$ a more universal behaviour is found when $T_c(\zeta)$ is plotted against $\rho_s(\zeta)$. This is shown by the solid curve in Fig. 1. It exhibits a canonical negative curvature similar to the data but with some systematic deviations. What complicates this issue is the fact that $T_c$ is governed both by the magnitude of the maximum SC gap, $\Delta_o$, and by the presence of the pseudogap. We resolve this by recognising that the ratio $T_c/\Delta_o$ is affected just by the pseudogap and thus we recast the Uemura plot as $T_c/\Delta_o$ versus $\rho_s(0)$. Note that $\Delta_o$ is the *spectral gap*, the magnitude of which remains unchanged by the pseudogap [14]. Our next goal, then, is to determine the ratio $T_c/\Delta_o$.

Fig. 4 shows the doping dependence of $T_c$ and $\Delta_0$. For Y-123 and La-214 $\Delta_0$ is determined from specific heat studies [11,17] using the linear slope in $\gamma(T)$ at low T:

$$\gamma(T) - \gamma(0) = 3.14 \gamma_n k_B T / \Delta_0, \qquad (3)$$

while the $\Delta_0$ values for Bi-2212 in Fig. 4 are obtained from $B_{1g}$ Raman scattering [20]. The general picture that emerges is that for $p > p_{crit}$ the ratio $T_c/\Delta_0$ remains constant and only begins to decrease with the opening of the pseudogap as $p$ falls below $p_{crit}$. This is precisely what is predicted by our model calculation. For convenience, we fit $\Delta_0$ to the form $\Delta_0(p) = \Delta_0(0) \tanh[\alpha(0.27-p)]$ (see solid curves in Fig. 4). Using this parameterisation we present a modified Uemura plot in Fig. 5 of the form $T_c/\Delta_0$ versus $\rho_s(0)$ for the three HTS materials.

The plots in Fig. 5 each display three regions. There is good agreement between data and calculation (solid red curve) from about 1/8[th] doping to critical doping. This confirms that it is the pseudogap that dominates $T_c$ and $\rho_s(0)$ here. The second region, $p > p_{crit}$, appears to be non-universal as noted above. In the third region, $p < 1/8$, $\rho_s(0)$ is higher than expected. We suggest this is due to the abrupt partial filling of the pseudogap, observed in this region from either the entropy [16] or the Knight shift [21]. Singer and Imai [22] have also recently drawn attention to this effect which, we think, may be associated with incipient stripe instabilities breaking up the pseudogap state. We have modelled this gap-filling using a triangular gap, $N(E) = N_1 + (N_0 - N_1) \times |E-E_F|/E_g$ with a finite DOS, $N_1$, at $E_F$. We take $N_1$ to increase linearly with doping then fall abruptly to zero at $p \approx 0.12$, as observed [16]. The result, shown by the dashed red curve in Fig. 5(a), follows the data well enough to illustrate that the canonical pseudogap behaviour is present for all $p < p_{crit}$, including $p < 1/8$.

We return now to the empirical relation, shown in Fig. 2, between $\lambda_0^{-2}$ and $S(T_c)/T_c$. This can be understood if $\rho_s(0)$ is expressed in terms of Fermi surface parameters using [23]

$$\rho_s(0) = \mu_o e^2 <v_x^2 N(E)> \tag{4}$$

where $v_x$ is the Fermi velocity projected in the $x$-direction and the average is taken over an energy shell $E_F \pm \Delta_0$. If $v_x^2$ is only weakly p-dependent (and we ignore its $k$-dependence) then, from eq. (4), $\rho_s(0) \propto <N(E)> \propto <\gamma>$ where $\gamma \equiv C_p/T$ is the specific heat coefficient. Now since $\gamma = \partial S/\partial T$ then $<\gamma> = (S/T)_{Tc}$. Thus, under these assumptions, $\rho_s(0) \propto (S/T)_{Tc}$ and the $p$-dependence of $\lambda_0^{-2} \times T_c$ is the same as that of $S(T_c)$, as demonstrated in Fig. 2. Taking $v_F^2 = 2 v_x^2$ where $v_F$ is the Fermi velocity, we have in detail:

$$\lambda_0^{-2} \times T_c = 3 \mu_o e^2 v_F^2 / (\pi^2 k_B 2V_a) \times S(T_c)/R \tag{5}$$

where $V_a$ is the volume of the unit atomic cell. For Y-123 Fig. 2 shows $T_c \times \lambda_0^{-2} = 3.7 \times 10^{16} \times S(T_c)/R$ in MKS units which, using eq. (5), yields a Fermi velocity $v_F = 1.35 \times 10^5$ m/s, in good agreement with other estimates [8]. Using the relation $\xi_0 = \hbar v_F/\pi\Delta_0$ and taking for $\Delta_0$ the angular average $\Delta_0/\sqrt{2}$ with $\Delta_0=270$ K at $p = p_{crit}$ [11] one finds $\xi_0 = 1.7$ nm, as observed [24].

Alternatively, the electronic entropy may be calculated from [14]

$$S^{el}(T) = -2k_B \int [f\ln f + (1-f)\ln(1-f)] N(E) dE , \tag{6}$$

where $f(E/k_BT)$ is the Fermi function. Using the NS DOS in eq. (1) we find that $S(T_c)/T_c$ has almost the same dependence on $\zeta \equiv E_g/(2.397 k_B T_c)$ as does $\rho_s(0)$ determined from eq. (2), in agreement with our empirical result shown in Fig. 2.

Another important observation shown in Fig. 2 is the numerical agreement between $S(T_c)/R$ and the critical concentration, $x_{crit}$, of Zn for the suppression of superconductivity (data from refs. 24-26). For a d-wave order parameter, and under the assumption that Zn is a unitary scatterer, $x_{crit} = 1.3 <N(E)> \Delta_0$ [25,26]. Since $<N(E)> \Delta_0$ is the pair density this implies that SC is destroyed when the density of unitary scatterers approximately equals the SC pair density i.e. each scatterer breaks one pair. We therefore conclude that $S(T_c)/R$ equals the density of pairs at $T=0$ in the Zn-free material, as expected if the pseudogap reflects a loss of normal-state spectral weight. If on the other hand $T_c$ and $S(T_c)$ were limited by thermally-induced SC fluctuations then $S(T_c)/R <<$ the density of pairs at $T=0$ and would not equal $x_{crit}$ [23]. The results of Fig. 2 are thus summarised by

$$\lambda_0^{-2} \times T_c \propto S(T_c)/R = x_{crit} . \tag{7}$$

A central conclusion here is that the doping-dependent magnitude of $\rho_s$ not only correlates with, but may also be estimated *numerically* from $S/T_c$. This ties the *p*-dependence of $\rho_s$ closely to that of the pseudogap and, as noted, provides a link between a ground-state SC property, $\rho_s(0)$, and a finite-temperature normal-state property, $S(T_c)$. Such a link naturally arises from the presence of a competing normal-state correlation together with the inference that $E_g$ is essentially *T*-independent. In the precursor pairing model of Levin and coworkers [6] a two-gap ansatz of the form

$$\Delta(\boldsymbol{k})^2 = \Delta'(\boldsymbol{k})^2 + E_g(\boldsymbol{k})^2 \qquad (8)$$

is employed (see also ref. [14]). While $E_g$ in this model may be relatively *T*-independent above $T_c$, its magnitude below $T_c$ falls to zero as $T\rightarrow 0$. In such a case $S(T_c)/T_c$ is diminished by precursor pairing but $\rho_s(T=0)$ is not. There is thus no reason for a direct correspondence between these two quantities, although quantitative details remain to be addressed. We believe that the observed relationship between $\rho_s$ and $S(T_c)/T_c$ implies that $E_g$ remains finite at $T=0$ i.e. the pseudogap is a ground-state correlation which coexists with SC even at $T = 0$.

In conclusion, we have shown that the identical doping dependence and observed simple numerical relations between $T_c\times\rho_s$, $S(T_c)$ and $x_{crit}$ place strong constraints on the nature of the normal-state pseudogap. A Fermi surface model of the pseudogap which successfully accounts for these correlations predicts a *sub-linear* dependence of $T_c$ vs $\rho_s$ over the entire underdoped region which agrees better with experiment than the usually-assumed linear dependence. With the opening of the pseudogap at $p_{crit}$ the superfluid density, $\rho_s$, falls rapidly at first while $T_c/\Delta_0$ falls at first slowly – the canonical variation expected if pseudogap correlations coexist with superconducting correlations in the ground state.

**Figure Captions**

Fig. 1. $T_c$ plotted as a function of superfluid density, $\lambda_0^{-2}$, for La-214 and Y:Ca-123. The arrow indicates increasing doping. Open data points denote 1/8th, optimal and critical doping. The dashed straight line shows the Uemura relation $T_c \propto \rho_s(0)$ and the solid curve is the calculated behaviour when a competing NS correlation opens up a gap in the DOS.

Fig. 2. The doping dependence of the critical concentration, $x_{crit}$, of Zn for suppressing superconductivity (red), of the electronic entropy, $S(T_c)/k_B$ (blue), and of $T_c \times \lambda_0^{-2}$ (black) for Y-123, Bi-2212 and La-214. For Bi-2212 the impurity scatterer is Co.

Fig. 3. The calculated normalised reduction in superfluid density and $T_c$ arising from a "triangular" gap in the NS DOS. $E_g$ is the temperature-independent NS gap.

Fig. 4. The doping dependence of $T_c$ (blue) and $\Delta_0$ (red) for Y-123, Bi-2212 and La-214. The solid curves are fits as described in the text.

Fig. 5. Modified Uemura plots in the form of $T_c/\Delta_0$ plotted as a function of $\lambda_0^{-2}$. The red solid curves are the calculated behaviour within the pseudogap model. The dashed red curve shows the gap-filling model calculation. The arrow indicates increasing doping.

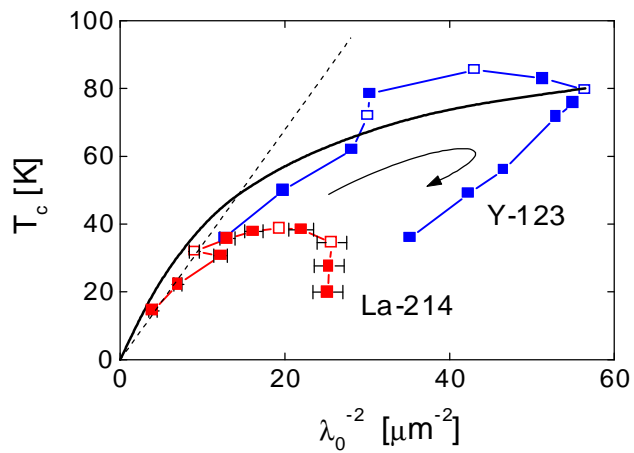
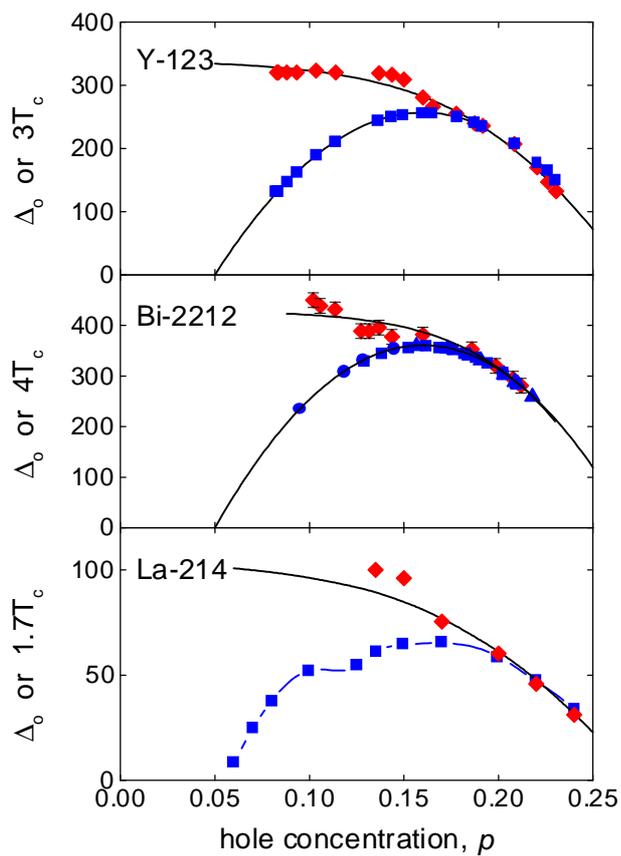
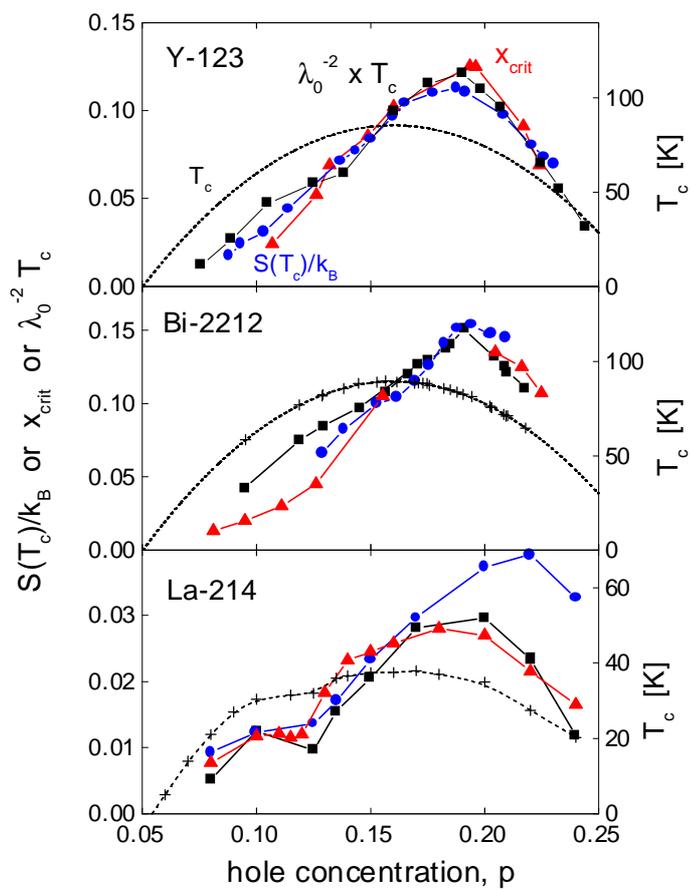
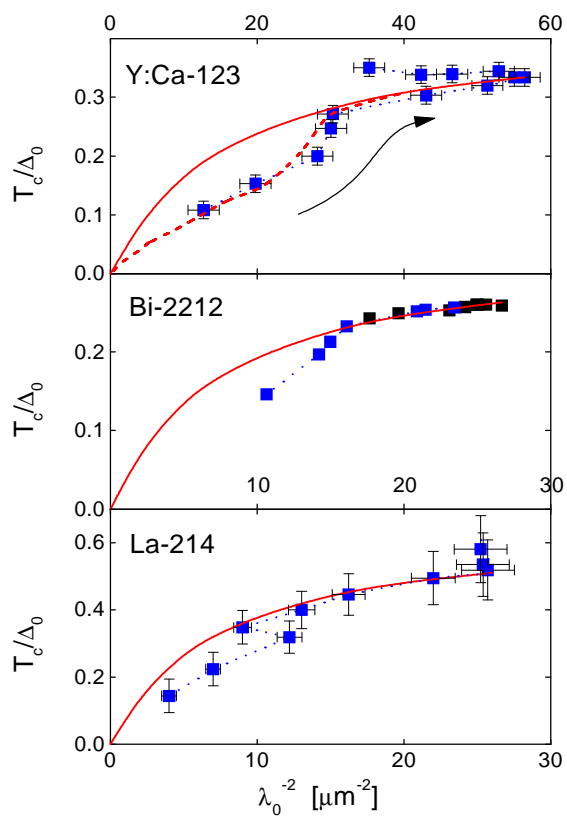
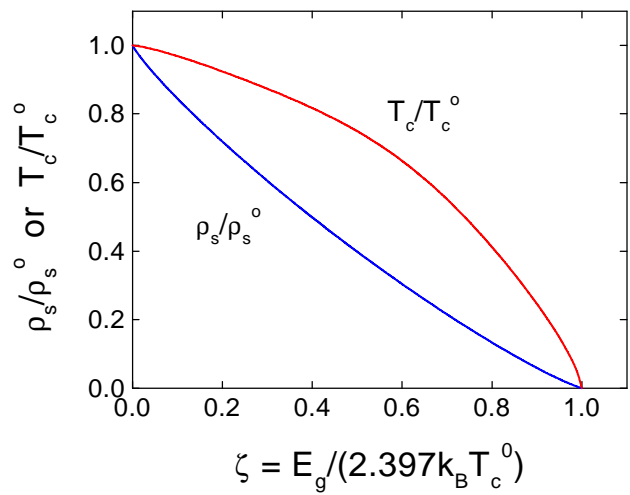